\documentclass[aps,prx,reprint,twocolumn,showpacs,superscriptaddress,a4paper,floatfix]{revtex4-1} 
\usepackage{comment}
\usepackage{hyperref}
\usepackage{graphicx}%
\usepackage{dcolumn}%
\usepackage{bm}%
\usepackage{latexsym}
\usepackage{amsmath}
\usepackage{natbib}
\usepackage{color}

\def\be{\begin{equation}}
\def\ee{\end{equation}}
\def\bea{\begin{eqnarray}}
\def\eea{\end{eqnarray}}

\allowdisplaybreaks
\begin{document}

\title{Non-linear response in the cumulant expansion for core hole photoemission}

\author{Marilena Tzavala}
\email[]{}
\affiliation{Department of Physics, University of Washington, Seattle, WA 98195-1560, USA}
\affiliation{Laboratoire des Solides Irradi\'es, \'Ecole Polytechnique, CNRS, CEA/DRF/IRAMIS, Institut Polytechnique de Paris, F-91128 Palaiseau, France}
\affiliation{European Theoretical Spectroscopy Facility (ETSF)}
\author{Joshua J. Kas} 
\affiliation{Department of Physics, University of Washington, Seattle, WA 98195-1560, USA}
\affiliation{European Theoretical Spectroscopy Facility (ETSF)}
\author{Lucia Reining}
\affiliation{Laboratoire des Solides Irradi\'es, \'Ecole Polytechnique, CNRS, CEA/DRF/IRAMIS, Institut Polytechnique de Paris, F-91128 Palaiseau, France}
\affiliation{European Theoretical Spectroscopy Facility (ETSF)}
\email[]{lucia.reining@polytechnique.fr}
\author{John J. Rehr}
\affiliation{Department of Physics, University of Washington, Seattle, WA 98195-1560, USA}
\affiliation{European Theoretical Spectroscopy Facility (ETSF)}

\date{\today}

\begin{abstract}
Most currently used approximations for the one-particle Green's function $G$ in the framework of many-body perturbation theory, such as Hedin's GW approximation or the cumulant GW+C approach, are based on a linear response approximation for the screened interaction $W$. The extent to which such a hypothesis is valid and ways to go beyond have been explored only very little.
Here we show how to derive a cumulant Green's function  beyond  linear-response from the equation of motion of the Green's function in a functional derivative formulation. The results can be written in a compact form, which opens the possibility to calculate the corrections in a first principles framework using time-dependent density functional theory. In order to illustrate the potential importance of the corrections, numerical results are presented for a model system with a core level and two valence orbitals.  
\end{abstract}

\maketitle
\section{Introduction}
Core-level x-ray photoemission spectroscopy (XPS) is a sensitive probe of correlation properties in condensed matter~\cite{Hedin_1999}. At high photon energies, where extrinsic and interference effects due to scattering of the outgoing photoelectron can be neglected, the XPS photocurrent is directly related to the spectral function associated with the core-hole Green's function. The importance of many-body effects for  spectra related to core excitations has  long been recognized, and the corresponding research has a rich history, going back at least to the 1920's~\cite{wentzel1925a,wentzel1925b,PhysRev.49.1}. Of particular interest are  edge singularities and asymmetric line shapes. These could be explained by the coupling between the core hole and the valence electrons using studies based on  model Hamiltonians, e.g., in the seminal works of Mahan, Nozi\`eres and De Dominicis, and coworkers~\cite{PhysRevB.25.5021,PhysRev.178.1072,PhysRev.178.1084,PhysRev.178.1097,Hedin_1999,Andersen_2001}. Several exact and approximate results were obtained, for example assuming the interaction potential is separable~\cite{PhysRevB.25.5021}. Another commonly used assumption is the absence of interaction between the valence electrons themselves, which implies that their scattering from the core hole potential leads to excitation of independent electron-hole pairs. In 
{the same} framework, the core-hole problem was treated by solving two coupled Bethe-Salpeter equations for core and valence electrons to lowest order in a parquet approximation~\cite{PhysRev.178.1072}, to which self-consistency in self-energy and vertex were added in~\cite{PhysRev.178.1084}. In the subsequent paper of the same series~\cite{PhysRev.178.1097}, an effective one-body approach was proposed, based on the calculation of the transient response to the sudden creation of a core hole. This was complemented by Langreth~\cite{PhysRev.182.973}, with a more compact derivation, and in \cite{PhysRevB.25.5150}, by an alternative approach based on a finite number of electrons in a box, which validated the approximation of a separable potential and added numerical illustrations. 

The picture of excitations created by the sudden appearance of the core hole is also naturally reflected in fermion-boson coupling Hamiltonians, where  the fermion refers to a deep core orbital, and the bosons are  electron-hole excitations, plasmons~\cite{ISI:A1969D745600003,PhysRevB.58.15565}, or phonons~\cite{ISI:A1975V957900001,Gunnarsson1994,ISI:000402871000002,PhysRev.145.602}. Remarkably, for a single fermion with linear coupling to a dispersing boson, the model can be solved exactly~\cite{Langreth-quasiboson1970,PhysRevB.17.929,Cini_1988}. The fermionic spectral function then consists of a quasi-particle peak followed by a Poisson series of satellites.  For increasing coupling strength the quasi-particle loses increasing weight to the satellites, and the envelope of the satellite spectrum becomes a Gaussian in shape. For this model, this exact solution is equivalent to the spectral function of the second-order (in the coupling constant) cumulant Green's function~\cite{Gunnarsson1994,Hedin_1999}. For more than one fermion energy and/or for higher-order coupling to bosons, the cumulant solution is not exact~\cite{Gunnarsson1994}; nevertheless since it contains the essential physics of the system responding by bosonic excitations to the the creation of an additional electron or hole, it is often a good approximation that is widely used, in particular for the core hole problem~\cite{Chang1972,Chang1973,ISI:A1980KP06500028,ISI:A1980JV80200003,huefner-pe,DeGroot2008,mahan-mpp}.

Of course the electron-boson picture is just another way to look at the old problem of many-body effects in core spectroscopy, but it highlights  three questions that one might ask, namely: \textit{i) How well does a Hamiltonian with linear electron-boson coupling describe the real problem?; ii) Which excitations should be contained in the boson?}; and  \textit{iii) Is the second-order cumulant solution good enough for the real application, or if not, how can one go beyond?} Heretofore these questions have not found a definite answer, not least because the domain of application of the cumulant approach in condensed matter is very wide, including for example coupling to phonons~\cite{ISI:000402871000002,PhysRevB.97.115145}, or valence electron spectroscopy~\cite{Aryasetiawan-cumulant1996,cumulant-Ferdi-Al-experiment1999,Kheifets2003,Guzzo-prl2011,Gatti2013,cumulant-Steven-2013,Guzzo2014,zhou-jcp2015,Lischner2015,Nakamura2016,Gumhalter2016,Caruso2016}.  

Analogous questions arise in an at first sight, different framework which is the calculation of Green's functions from a Dyson equation, where interaction effects are contained in the self-energy, the integral kernel of the equation~\cite{Fett71}. This is the framework now commonly adopted in first principles calculations. Currently the most widely used approximation is Hedin's GW approximation (GWA)~\cite{Hedin1965}, where the self-energy is approximated as the product of the one-electron Green's function $G$ and the screened Coulomb interaction $W$ to first order. Indeed, one can view the GW approximation as an approximate solution of the electron-boson problem \cite{Hedin_1999}, where the bosonic excitations are the excitations contained in $W$. Also in this case, since the coupling is only linear, the boson contains only a selection of excitations (spin flip excitations, for example, are neglected), and the problem is solved only approximately. While the GWA  has been very successful for the calculation of quasi-particle energies, it typically gives a poor description of satellite spectra~\cite{Martin2016}.  This shortcoming can be traced  to the third of the above questions - the approximation used in the solution of the Hamiltonian problem - since this approximation leads to problems such as the appearance of a spurious plasmaron solution. This can be seen in the GW solution of the exactly solvable  model~\cite{Hedin_1999}. Changing from the GWA to a cumulant Green's function leads to a significant improvement without much additional effort since the same boson appears through $W$ but the electron-boson problem is solved more accurately. The GW plus cumulant (GW+C) approach for the one-electron Green's function~\cite{Aryasetiawan-cumulant1996,Guzzo-prl2011} has recently become a  popular first principles method to describe spectral functions, including satellite series. Here, the satellites are mostly due to plasmons~\cite{zhou-jcp2015}, the dominant excitations in $W$. 

The success of the GW and GW+C approximations may seem surprising in view of the relative simplicity of these approximations being linear in $G$ and $W$. While it is physically reasonable, e.g., in many simple semiconductors, that plasmonic excitations should represent the dominant boson, it is less obvious that the physics is sufficiently described by \textit{linear} response. This is especially evident in systems with few electrons, or when the removal or addition orbital is very localized, where one might expect that response terms of higher order should play a role. These terms are not contained in the electron-boson model with linear coupling, because the boson itself is fixed and does not respond to the excitation; analogously, they are not contained in GW nor GW+C 
because $W$ is calculated on the ground state, in presence of the original $N$ electrons. The fact that non-linear screening should be present and should show up in a cumulant solution has been pointed out early on by Mahan~\cite{PhysRevB.25.5021}, who calculated the leading correction to the standard second-order (in the coupling constant) linear-response cumulant solution. In that work the  valence electrons that respond to a core hole excitation are described by an independent-electron picture, and even so, going to yet higher orders turned out to be too complicated. While this pioneering work, as well as the seminal solution of Nozi\`eres and De Dominicis~\cite{PhysRev.178.1097}, trace a way to go towards the inclusion of non-linear screening effects, these model approaches are not directly transferable to first principles calculations, for several reasons: i) because of the approximations involved from the very beginning on the interaction potential; ii) because of the absence of interaction between valence electrons, which would lead to a poor description of screening with the absence of plasmons; and iii) because it is not clear how an expansion that is order-by-order concerning the response functions (i.e., linear response, second order response, etc.) would converge. Progress in several directions has been made, such as a better description of screening~\cite{ISI:A1976BQ07000036,PhysRevB.26.2772}, higher-order cumulant solutions of the electron-boson model, such as in \cite{Gunnarsson1994,Gumhalter2005,ISI:000408102500051} the derivation of higher order correlation functions from the Mahan-Nozi\`eres-DeDominicis (MND) or electron-boson Hamiltonians~\cite{PhysRevB.72.235110},
or more recent progress with the description of non-linear electron-boson couplings in Ref.\ \cite{PhysRevB.98.075105} and in the equation-of-motion, coupled-cluster  method\cite{Rehr-JChemPhysinpress2020}. However, there is still a gap to be filled concerning the \textit{ab initio} calculation of spectral functions. 

The present work aims at developing a robust, first-principles derivation of non-linear screening effects in a cumulant Green's function. While we do not claim to invent new physics here, our derivation has the advantage of being  compact, and situated within the framework now commonly used in the \textit{ab initio} community. Our approach does not depend on a contact or separability approximation on the interaction potential. Moreover, the approach highlights the underlying physics, which is crucial if one wishes to understand when non-linear effects are  important, and where should be the limits of their applicability. Finally, it allows us to propose a way to put the equations into practice, by combining the many-body perturbation formalism with time-dependent density functional theory (TDDFT)~\cite{Runge1984}. In this way, an order-by-order expansion of the response is avoided, and the problem of calculating the effects of screening on the Green's function is separated from that of the calculation of the screening itself. 
Contrary to numerous previous works on core spectroscopy, we are not so much interested in asymmetry of line shapes but rather, in the satellites, for which the standard second-order linear response cumulant approach is now a well established first principles approach and convenient starting point. 

The paper is organized as follows: The background concerning the formalism and the GW approximation is contained in Sec. \ref{sec:background}. Next, we derive the core-hole cumulant Green's function in its various approximations in Sec. \ref{sec:cumulant}, and we analyze the results in Sec. \ref{sec:analysis}. Numerical results for an illustrative model are given in Sec. \ref{sec:model}, and a short conclusion is given in Sec. \ref{sec:conclusions}.

\section{Background}
\label{sec:background}

Beginning with its equation of motion following the approach of Martin and Schwinger \cite{PhysRev.115.1342}, 
the one-body Green's function can be described by a functional differential equation
which is often referred to as the Kadanoff-Baym equation~\cite{PhysRev.124.287}. The equation was initially derived for temperature dependent, non-equilibrium quantum systems. However, 
it can also be used  to create the diagrams describing equilibrium and/or zero-temperature physics~\cite{Martin2016}, as we will do here.
In this formalism, the fully interacting propagator $G$ is given by
\begin{align}
G(12)=G_0(12)&-iG_0(1\bar{1})v(\bar 1\bar 3)G(\bar 3\bar 3^+)G(\bar 12)\nonumber\\
&+iG_0(1\bar 1) v(\bar 1\bar 3)\frac{\delta G(\bar 12)}{\delta u(\bar 3^+)},\label{eq:KBE_G0U}
\end{align}
where $G_0$ is the non-interacting propagator in the presence of $u$,
$v(12)=\delta(t_1-t_2)/{|{\bf r}_1-{\bf r}_2|}$ the Coulomb interaction, and $u(3)$ is a
local, time-dependent external perturbing potential that simulates interaction effects due to the propagation of particles; it will be taken to zero at the end of the calculation. Here and below we employ the usual notation of an integer for a set of space, spin and time variables ($1\rightarrow(r_1,\sigma_1,t_1)$), and bars for variables that are integrated over: $f(\bar 1)g(\bar 1)\equiv\int d1 f(1)g(1) $.  All quantities are functionals of $u$. The classical Hartree term (the second term on the right) depends on the interacting density as given by the diagonal part of the propagator $n(1)=-iG(11^+)$, where $1^+\equiv \lim_{\eta\to 0}({\bf r}_1,\sigma_1,t_1+\eta)$, $\eta>0$.  The last term contains the functional derivative of $G$, which accounts for the exchange and correlation effects. This term turns Eq.\ (\ref{eq:KBE_G0U}) into a first-order non-linear functional differential equation with respect to $u$. 
 Introducing the total classical potential $u^H$, one obtains a set of coupled equations for $u^H$ and $G$,
\begin{align}
&u^H(1)=u(1)+v(1\bar 3)n(\bar 3),\\
&G^H(12) = G_0(12)+G_0(1\bar 4)
{u^H(\bar 4)}G^H(\bar 4,2),\\
&G(12)=G^H(12)+G^H(1\bar{1})v(\bar 1\bar 3)\frac{\delta G(\bar 12)}{\delta u(\bar 3^+)}.
\label{eq:original-kbe}
\end{align}
In extended systems, screening plays an important role. For this reason, is is often convenient to rewrite the functional derivative  using the chain rule with respect to the classical potential,
\begin{equation}
    G(12)=G^H(12)+iG^H(1\bar{1})W(\bar 1\bar 4;u)\frac{\delta G(\bar 12)}{\delta u^H(\bar 4^+)},
    \label{eq:screened-kbe}
\end{equation}
where
\begin{equation}
    W( 1 4;u)\equiv v(1\bar 3)\frac{\delta u^H(4)}{\delta u(\bar 3)}.
\end{equation}
Note that here the screened potential $W$ is a functional of the external potential, such that the equation remains exact.

Often, the dependence of $W$ on $u$ is neglected. 
The solution of this linear-response version of \eqref{eq:screened-kbe} in a simple model has been discussed in~\cite{Lani_2012}. In general, however, even in the linear response approximation, the equation cannot be solved exactly. Therefore, the functional derivative on the right side is usually approximated such that the limit $u\to 0$ can be taken directly. 
One of the  most widely used approximations for the Green's function is Hedin's GW approach~\cite{Hedin1965}, which is obtained by  approximating the functional derivative,
\begin{equation}
  \frac{\delta G( 12)}{\delta u^H( 4)}\approx G(14)G(42).
  \label{eq:der-gw}
\end{equation}
Then the $u\to 0$ limit can be taken, and one has a Dyson equation for $G$,
\begin{equation}
G(12) = G^H(12) + G^H(1\bar 3)\Sigma_{\rm xc}^{GW}(\bar 3\bar 4)G(\bar 42)
\end{equation}
with 
\begin{equation}
    \Sigma_{\rm xc}^{GW}(14)  \equiv iG( 14)W(14;u\to 0).
\end{equation}
The GW approximation (GWA) has been very successful for the calculation of quasi-particle energies. However, a major shortcoming is its poor treatment of the satellite part of electron addition or removal spectra~\cite{Aryasetiawan-cumulant1996,Guzzo-prl2011,zhou-jcp2015,Martin2016}.  Calculations based on a cumulant Green's function which yields a much much better satellite spectrum have been found to be an advantageous alternative. The cumulant approach, which avoids the approximation in Eq.(\ref{eq:der-gw}) that leads to the GW self-energy, will be discussed in the following section.

\section{Derivation of the core-hole cumulant }
\label{sec:cumulant}

In order to describe the photoemssion spectra from core states, we need 
the projection of the Green's function on the core orbital $G_{cc}$. Its associated spectral function simulates the core photoemission spectrum, especially at high photon energies, where the photoelectron and associated effects of extrinsic losses can be ignored. Thus here we  make the approximation that the core hole is decoupled from all other orbitals, except for the screening of the interaction, which is due to the valence electrons. The {\it decoupling approximation} is physically reasonable for the case of localized electrons, such as a  deep-core state, which has little overlap with the valence electrons. For this case, we may suppose that the interacting and the Hartree Green's functions $G$ and $G^H$ do not have matrix elements linking the core and a valence state, so Eq.\ (\ref{eq:original-kbe}) written in a basis of single-particle orbitals reads
\begin{align}
G_{cc}(t_1t_2)&=G_{cc}^{H}(t_1t_2) + \nonumber\\
&+iG_{cc}^{H}(t_1\bar t_3)\sum_{kl}v_{cckl}\frac{\delta G_{cc}(\bar t_3t_2)}{\delta u_{kl}(\bar t_3^+)},\label{eq:KBE_GHU}
\end{align}
where the matrix elements of the Coulomb interaction are defined as 
$$v_{cckl}=\int d{\bf r}\phi_l^{\star}({\bf r})\phi_k({\bf r})\int d{\bf r}'{|\phi_c({\bf r}')|^2}/{|{\bf r}-{\bf r}'|}.$$
In principle, this equation is understood to be on a contour, since in presence of the time-dependent external potential the system is out of equilibrium~\cite{non-equibrium-robert}. As an additional approximation, we only retain the contributions that correspond to the propagation of a hole, i.e. $t_2 > \bar t_3> t_1$. This allows us to make the ansatz that 
 the interacting propagator is proportional to $G^H$, i.e.
\begin{equation}
G_{cc}(t_1t_2)= G_{cc}^{H}(t_1t_2) F(t_1t_2). 
\end{equation}
With the identity
$
\frac{\delta G}{\delta U}=\frac{\delta G^H}{\delta U}F+G^H\frac{\delta F}{\delta U},
$
we obtain
\begin{align}
  F(t_1t_2)&=1  +i\frac{G_{cc}^{H}(t_1\bar t_1)}{G_{cc}^{H}(t_1t_2)}\times \nonumber\\
&\times \sum_{kl}v_{cckl}\Big [\frac{\delta G^H_{cc}(\bar t_1t_2)}{\delta u_{kl}(\bar t_1^+)}F(\bar t_1t_2) \nonumber\\
&+ G_{cc}^H(\bar t_1t_2)\frac{\delta F(\bar t_1t_2)}{\delta u_{kl}(\bar t_1^+)}\Big ].\label{eq:KBE_GHU-2}
\end{align}
As in the derivation of the GWA, we can now use the chain rule with the total classical potential, and we again suppose that  $G^{H}$ has no off-diagonal elements linking core and valence single-particle states. This leads to
\begin{align}
  F(t_1t_2)&=1 +i\frac{G_{cc}^{H}(t_1\bar t_1)}{G_{cc}^{H}(t_1t_2)} \times\nonumber\\
&\times \Big [W_{c}(\bar t_1^+\bar t_4;u)G^H_{cc}(\bar t_1\bar t_4)G_{cc}^H(\bar t_4t_2)F(\bar t_1t_2) \nonumber\\
&+\sum_{kl}v_{cckl} G_{cc}^H(\bar t_1,t_2)\frac{\delta F(\bar t_1t_2)}{\delta u_{kl}(\bar t_1^+)}\Big ],\label{eq:KBE_GHU-3}
\end{align}
with the screened interaction $W_c\equiv W_{cccc}$, where 
\begin{eqnarray}
&W_c\equiv W_{cccc}(t_1^+t_4)&=v_{cccc}\delta(t_4t_1^+)\nonumber\\
&&+\sum_{klk'l'}v_{cckl}v_{cck'l'}\frac{\delta n_{kl}(t_4)}{\delta u_{k'l'}(t_1^+)}.
\label{eq:screening}
\end{eqnarray}
At this stage $W_{c}(t_1t_4;u)$ still depends on the potential $u$, which has not yet been set to zero. 
Also note that \eqref{eq:screening} suggests a response of the density; however, one has to keep in mind that here we are not in a retarded framework, but  only keep parts corresponding to the time ordering defined above.

With the constraint  $t_2>t_1$ imposed on the time orderings, $G_{cc}^H =i e^{-i\varepsilon_c^H(t_1-t_2)}$ factorizes, and the factors cancel. This yields
\begin{align}
  F(t_1t_2)&=1 -i 
 \int_{t_1}^{\infty}d\bar t_1\int_{\bar t_1}^{t_2} d\tau\,W_{c}(\bar t_1^+\tau;u)F(\bar t_1t_2) \nonumber\\
&-\sum_{kl}v_{cckl} \int_{t_1}^{t_2} d\bar t_1\,\frac{\delta F(\bar t_1t_2)}{\delta u_{kl}(\bar t_1^+)}.\label{eq:KBE_GHU-4}
\end{align}
Next, we define $F$ in cumulant form 
\begin{equation}
    F(t_1t_2)\equiv e^{C(t_1t_2)}
\end{equation}
 and take the derivative of Eq.\ (\ref{eq:KBE_GHU-4}) with respect to $t_1$. This yields a differential equation for the cumulant function $C$
\begin{align}
\partial_{t_1}C(t_1t_2)&=i\int_{t_1}^{t_2} d\tau W_c(t_1\tau;u)\nonumber\\
&+\sum_{kl}v_{cckl}\frac{\delta C(t_1t_2)}{\delta u_{kl}(t_1^+)}.
\end{align}
With the boundary condition $C(t_2t_2)=0$, the integral equation for $C$ is given by
\begin{align}
C(t_1t_2)&=-i\int_{t_1}^{t_2}d\tau'\int_{\tau'}^{t_2} d\tau W_c(\tau'\tau;u)\nonumber\\
&-\sum_{kl}v_{cckl}\int_{t_1}^{t_2}d\tau'\frac{\delta C(\tau't_2)}{\delta u_{kl}(\tau'^+)}.
\label{eq:C}
\end{align}
The last term of this expression is commonly neglected, so one can   directly set the external potential to zero. This yields the widely used cumulant in the linear response approximation
\begin{align}
C^0(t_1t_2)&=
-i\int_{t_1}^{t_2} d\tau'\int_{\tau'}^{t_2} d\tau W_c(\tau'\tau;u) \nonumber\\
&=-i\int_{t_1}^{t_2} d\tau\int_{t_1}^{\tau} d\tau' W_c(\tau'\tau;u).
\label{eq:C0}
\end{align}
For the core hole, together with Eq.\ (\ref{eq:screening}) the expression can be interpreted as the integral over the variation of the  Coulomb potential due the valence density at time $\tau$, induced in linear response of the system by the creation of a hole at time $\tau'$. The response is causal with $\tau > \tau'$, and the process is integrated over the interval $(t_1,t_2)$. Implicit in the  evaluation of the cumulant $C^0$  is that only the positive frequency components of the Fourier transform $W_c(\omega)$, corresponding to lossy excitations, are present in the linear response approximation~\cite{Langreth-quasiboson1970}. 

To go beyond linear response, one can iterate \eqref{eq:C}. The lowest order corrections stems from the  derivative of $C^0$, 
 taking into account that ${\delta W_c}/{\delta u}\neq 0$. This 
causes a second order response function to appear, in a contribution $C^1$ given by
\begin{eqnarray}
&C^1(t_1t_2)&=i\sum_{kl}v_{cckl}\times\nonumber\\
&&\times\int_{t_1}^{t_2} d\tau\int_{t_1}^{\tau} d\tau'\int_{t_1}^{\tau'} d\tau''\frac{\delta W_c(\tau'\tau;u)}{\delta u_{kl}(\tau''^+)}.
\label{eq:C1} 
\end{eqnarray}
This term contains the variation of the response of the valence density due to the density change caused by the linear response to the creation of the core hole. 

Higher order corrections involve higher order derivatives of $W_c$, and therefore higher order non-linear response functions. They should be evaluated from the density-density correlation functions in the ground state, i.e. before the removal of the core-electron.
The general form of the solution is thus given by 
\begin{equation}
C(t_1t_2)=\sum_{m=0}C^m(t_1t_2;u=0),
\end{equation}
where the  $m^{th}$ contribution $C^m$ is obtained recursively using the relation
\begin{eqnarray}
&C^{m+1}(t_1t_2;u=0)&=-\sum_{kl}v_{cckl}\times\nonumber\\
&&\left.\int_{t_1}^{t_2}d\tau\frac{\delta C^m(\tau t_2;u)}{\delta u_{kl}(\tau^+)}\right|_{u=0},\label{eq:recursive for C}
\end{eqnarray}
with $C^0$ given by Eq.\ (\ref{eq:C0}) in terms of $W_c$. 
The recursive formulation suggests that the series might in practice be truncated at a given order $n$.
Once all orders of the correction have been found, the limit of $u\to0$ is applied.
The full recursive solution implied by Eq.\ (\ref{eq:recursive for C}) is  the first main result of this paper. It may be seen as an extension of Mahan's approach in~\cite{PhysRevB.25.5021}, which contains the first non-linear response contribution. The main difference is a more general formulation such that i) a separable potential is not needed needed for the derivation: ii) the result is valid to infinite order in the response; and most importantly, iii) the valence electrons are in principle fully interacting.   As we will see below, the fact that the result is formulated in general terms of response functions rather than explicit sums over transitions, allows us to benefit from the use of TDDFT for an efficient approximation to the non-linear response.



\section{Analysis}
\label{sec:analysis}

\subsection{Effective interaction }

The cumulant $C^0(tt')$ in Eq.\ (\ref{eq:C0}) and consequently  the higher order terms, are  double integrals of a two-time function over time. We can therefore  express the cumulant in terms of a new function $w(\tau\tau')\equiv\sum_{m=0} w^m(\tau\tau')$, where $w^0(\tau\tau')=W_c(\tau'\tau)$ and $w^m$ stands for the order $m$ correction in $W_c$. Each cumulant term is related to $w$ through the expression 
\begin{equation}
C^m(t_1t_2)=-i\int_{t_1}^{t_2} d\tau\int_{t_1}^{\tau}d\tau' w^m(\tau\tau').
\end{equation}
A recursive relation, similar to Eq.\ (\ref{eq:recursive for C}), holds between the different orders of $w$,
\begin{equation}
 w^{m+1}(\tau\tau')=-\sum_{kl}v_{cckl}\int_{\tau'}^{\tau} d\tau''\frac{\delta w^m(\tau\tau''; u)}{\delta u_{kl}(\tau'^+)},\label{eq:recursive for w}
\end{equation}
where $\tau>\tau'$, and $\tau$ the time that refers to the variations of the density. The interaction $-v_{cckl}$ couples the core level from one side, to the variations of the external potential taken with valence levels on the other side. The negative sign can be understood in terms of the sign of a core-hole charge 
and $w^m$ can also be viewed in terms of the orders of an expansion to this core-hole potential. Therefore the full matrix $w(\tau \tau')=\sum_m^n w^m(\tau \tau')$ plays the role of an effective interaction, accounting for all orders of the density variations due to the propagation of a core-hole.
\subsection{Induced density variations}

In order to enable further interpretation and practical use, it is convenient to expand Eq.\ \eqref{eq:recursive for w} for $m>0$,
\begin{eqnarray}
 &&w^{m}(\tau\tau')=(-1)^m\sum_{k_1l_1,...k_ml_m}v_{cck_1l_1}...v_{cck_ml_m}\times\nonumber\\
 &&\int_{\tau'}^{\tau}..\int_{\tau'_{m-1}}^{\tau} d\tau_1..d\tau_m\frac{\delta^m w^0(\tau\tau_m; u)}{\delta u_{k_1l_1}(\tau'^+)\delta u_{k_2l_2}(\tau_1^+)...\delta u_{k_ml_m}(\tau_{m-1}^+)}.\nonumber\\
 \label{eq:recursive for w-expand}
\end{eqnarray}
With $    w^0(\tau\tau')=W_c(\tau\tau')$ and \eqref{eq:screening}
the cumulant is given in terms of the induced valence density
{variations} as
\begin{eqnarray}
 &&   C(t_1t_2)=-iv_{cccc}(t_2-t_1)\nonumber\\
 &&\,\,\,\,\,\,\,\,\,\,+i\int_{t_1}^{t_2}d\tau\sum_{m=0}^{\infty}v_{cck_1l_1}...v_{cck_ml_m}v_{ccij}v_{ccrs}\nonumber\\
 &&\times \frac{(-1)^{(m+1)}}{(m+1)!}\int_{t_1}^{\tau}d\tau '\int_{t_1}^{\tau}...\int_{t_1}^{\tau} d\tau_1....d\tau_m\nonumber\\
 &&\,\,\,\,\,\times \frac{\delta^{(m+1)} n_{ij}(\tau)}{\delta u_{k_1l_1}(\tau')\delta u_{k_2l_2}(\tau_1)...\delta u_{k_ml_m}(\tau_{m-1})\delta u_{rs}(\tau_m)},
 \label{eq:cum-density}
\end{eqnarray}
where the factor $1/(m+1)!$ is due to the extension of the integration domain, and repeated indices are summed over.

The first term yields the exchange correction to the energy of the core level. 
The remainder can be summed: it is an expansion of the Coulomb potential acting on the core hole, created by the change in density $\Delta n$ at time $\tau$, which is in turn due to the potential $-v_{cc}({\bf r})\theta(t-t_1$ created by the switching on of a core hole at time $t_1$. This potential is then integrated in $\tau$ over the time of propagation of the core hole. The compact result can be written as
\begin{eqnarray}
   C(t_1t_2)&=&-iv_{cccc}(t_2-t_1)\nonumber\\
&&\,\,\,\,\,\,\,\,\,\,+i\int_{t_1}^{t_2}d\tau\,v_{ccij}\Delta \tilde n_{ij}(\tau).
\label{eq:final}
\end{eqnarray}
However, as pointed out earlier, $\Delta\tilde n$ is not equal to the causal response of the density. In particular, in order to extend the integration domains and obtain this compact result, the symmetry of the time-ordered $W$ has been used. In order to use the result, we therefore have to make an approximation.
One simple 
{possibility} is to use the real time-dependent induced density in Eq.~(\ref{eq:final}). However, this approach produces negative spectral weight on the high energy side of the quasiparticle (see Fig.~\ref{fig:model}). As an alternative approximation here we restrict $\Delta \tilde n_{ij}(\tau)$ to include only the positive frequency
components of the induced density, as the case in linear response.
This result has a
physical interpretation which moreover has a significant practical advantage: since the core orbital is kept fixed the perturbation is known and its effect can be calculated directly. For example, the  TDDFT in real space and time can be used, as proposed in~\cite{PhysRevB.94.035156} for the case of linear response. It is then not necessary to calculate the rather clumsy higher order response functions, which thereby overcomes the issues discussed in~\cite{PhysRevB.25.5021}. 

When the core hole is suddenly switched on, the system may react violently. However, an interacting system will eventually reach a new equilibrium,  given by the final state of the system with a static core hole. In this limit $\Delta \tilde n_{ij}(\tau)$ becomes independent of $\tau$. The simplest approximation is therefore a shift of the core level due to the fact that the valence density should be calculated in presence of the core hole. Since this approximation completely neglects dynamical effects, it cannot lead to satellites in the spectral function. In order to do better, one 
{could} apply the same reasoning starting from the next order, which means in practice, to evaluate explicitly the lowest order cumulant with the valence density calculated in presence of the core hole. Such an intuitive approach, which has been used successfully (see e.g.,~\cite{ISI:000235828300008,ISI:000229275600032,PhysRevB.70.195214}), 
{may be} justified by our derivation.

\subsection{Self-energy}
Let us now compare the above result with what one would obtain in the context of  a Dyson equation. For this case, we start again from the exact expression in Eq.\ (\eqref{eq:screened-kbe}) and then using $\delta G/\delta u^H = -G(\delta G^{-1}/\delta u^H)G$,  the exact Dyson equation becomes
\begin{eqnarray}
   && G(12)=G^H(12)\nonumber\\
   &&-iG^H(1\bar{1})W(\bar 1\bar 4;u)G(\bar 1\bar 5)\frac{\delta G^{-1}(\bar 5\bar 6)}{\delta u^H(\bar 4^+)}G(\bar 62),
    \label{eq:screened-kbe-inv}
\end{eqnarray}
which has a self-energy given by
\begin{eqnarray}
    \Sigma_{\rm xc}(16) &=& -i W(1\bar 4;u)G( 1\bar 5)\frac{\delta G^{-1}(\bar 56)}{\delta u^H(\bar 4^+)}\nonumber\\
    &=& i W(1\bar 4;u)G( 1\bar 5)\times\nonumber\\
    &\times&\Big(\delta(\bar 56)\delta(\bar 5\bar 4) + \frac{\delta \Sigma_{\rm xc}(\bar 56)}{\delta u^H(\bar 4^+)} \Big ).
    \label{eq:def-sigma}
    \end{eqnarray}
    Again taking the matrix element in the core orbital,  
    and using the relation  $G_{cc}( t_1t_5) = -iG_{cc}( t_1t_6)G_{cc}( t_6t_5)$, the core level self-energy is,
    \begin{eqnarray}
    \Sigma_{\rm xc}(t_1t_6)
    &=& i G_{cc}( t_1t_6)\Big (W_c(t_1t_6;u)\nonumber \\
    &-&iW_c(t_1\bar t_4;u)G_{cc}( t_6\bar t_5)\frac{\delta \Sigma_{\rm xc}(\bar t_5t_6)}{\delta u^H_{cc}(\bar t_4^+)}\Big )\nonumber \\
    &\equiv& i G_{cc}( t_1t_6) w^{\rm eff}(t_1t_6).
\end{eqnarray}
This self-energy has the same structure as the GWA, but instead of a linear response $W$, one has an effective interaction
\begin{eqnarray}
    && w^{\rm eff}(t_1t_6;u) \equiv W_c(t_1t_6;u) \nonumber\\
     &&\,\,\,\,\,\,\,\,\,\,+i W_c(t_1\bar t_4;u)G_{cc}( t_6\bar t_5)\frac{\delta \Sigma_{\rm xc}(\bar t_5t_6)}{\delta u^H_{cc}(\bar t_4^+)}\nonumber\\
     &=& W_c(t_1t_6;u) - W_c(t_1\bar t_4;u)\frac{\delta w^{\rm eff}(\bar t_5t_6;u)}{\delta u^H_{cc}(\bar t_4^+)}\nonumber\\
     &+& iW_c(t_1\bar t_4;u)G_{cc}( t_6\bar t_5) G_{cc}(\bar t_5\bar t_4^+)G_{cc}(\bar t_4^+t_6) w^{\rm eff}(\bar t_5t_6).\nonumber\\
     \label{eq:effective-full-w}
\end{eqnarray}
If the contribution in the last line is neglected, i.e., if one neglects the variation of the core hole Green's function,  $w^{\rm eff}=w$  is the effective interaction \eqref{eq:recursive for w} that appears in the cumulant. To lowest order neither the neglected term nor the functional derivative of $w^{\rm eff}$ contribute. This is also the reason why the GW self-energy and the lowest order cumulant have the same effective interaction, namely the linear response screened Coulomb interaction $W$. For the higher orders, however, it is important that the effective interactions are different, since $w$ has to be used in the cumulant and $w^{\rm eff}$ in the Dyson equation, while both are exact in principle within the approximations made here. For example, again making use of the product property of the core hole Green's function, the last term gives rise to a contribution $G_{cc}(t_1\bar t_5)W_c(t_1,\bar t_4^+)G_{cc}(\bar t_5\bar t_4^+)G_{cc}(\bar t_4^+t_6) W_c(\bar t_5t_6)$, which is neglected in the GWA, but already included in the cumulant Green's function  through the lowest order cumulant function.

Finally, the form $\Sigma_{\rm xc}=iGw^{\rm eff}$ is reminiscent the 
T-matrix formalism~\cite{Martin2016,PhysRevB.85.155131}.  Such T-matrix approximations can be derived by supposing that the self-energy consists of a Green's function and an effective interaction which, inserted into  Eq. (\ref{eq:def-sigma}), allows one to derive a Dyson equation for the effective interaction. However, for our present core-hole problem we have assumed that the core Green's function does not vary 
{whereas} derivatives of the interaction are taken into account. Instead, to obtain the scattering diagrams in the T-matrix approximations the Green's function is varied and the effective interaction is kept fixed~\cite{Martin2016,PhysRevB.85.155131}. This is more important in a situation of low density (partial filling), which is quite different from the core hole problem.

\section{Model calculations}
\label{sec:model}

As a concrete illustration of the approach presented here, we introduce a simple 3-state model, similar to that used by Lee, Gunnarsson and Hedin in~\cite{PhysRevB.60.8034}, which has also been used to treat charge-transfer satellites in x-ray spectra\cite{Klevak2014}.
This model system with two electrons, a core electron and a valence electron propagating in two atomic levels $a$, $b$, is described by the Hamiltonian
\begin{equation}
\hat{H}=\epsilon_0 \hat c^{\dagger}\hat c+\epsilon_a^0\hat n_a+\epsilon_b^0\hat n_b+U\hat n_h \hat n_a-t(\hat{c}_a^{\dagger}\hat c_b+\hat{c}_b^{\dagger}\hat c_a),\label{eq:initial hamiltonian}
\end{equation} 
where $\epsilon^0$ are atomic energies evaluated in the presence of the core-electron, and $U$ is the potential from the core-hole $\hat n_h=1-\hat n_c$ coupling only to one of the two levels, $a$. The valence levels belong to different atoms and this justifies the approximation to consider weak coupling to the level $b$. The hybridization between the levels $a$ and $b$ is represented by the interaction parameter $t$. This model aids a physical understanding of core-photoemission in molecules with flat valence bands, where charge transfer excitations between different atoms modify the spectrum. 

The initial state, where $n_h=0$, is described by the two-particle state
{$|\psi^{i}_0\rangle=\sin\phi|a\rangle|c\rangle+\cos\phi|b\rangle|c\rangle$}
that mixes core and valence levels, with
$\tan 2\phi={2t}/{\epsilon}$ and $\epsilon=\epsilon_a^0-\epsilon_b^0$. The energy of the initial state is given by $\epsilon^i_{1,0}=\frac{1}{2}(\epsilon_a^0+\epsilon_b^0)\pm\frac{1}{2}\sqrt{\epsilon^2+4t^2}$. 
The final states where $n_h=1$ have only one particle, and  are given by the single-particle wavefunctions
$|\Psi^f_1\rangle=\cos\theta|a\rangle-\sin\theta|b\rangle$ and
$|\Psi^f_2\rangle=\sin\theta|a\rangle+\cos\theta|b\rangle$, with
$\tan 2\theta={2t}/{(\epsilon+U)}$ and
$\epsilon^f_{1,2}=\frac{1}{2}(\epsilon_a^0+\epsilon_b^0+U)\pm\frac{1}{2}\sqrt{(\epsilon+U)^2+4t^2}$. 
In the model the core-hole potential couples only to the level $a$ and therefore the interaction with the time-dependent occupation of the level $a$ will appear. 

In order to obtain the cumulant solution for this model, we apply Eq.\ \eqref{eq:cum-density} 
\begin{align}
&C(t_1t_2)= {iUn_{a}^{0}(t_2-t_1)
+}iU^2\int_{t_1}^{t_2} d\tau\int_{t_1}^{\tau} d\tau'\frac{\delta n_a(\tau)}{\delta u_a(\tau')}_{|u_a=0}\nonumber\\
&+i\frac{U^3}{2}\int_{t_1}^{t_2} d\tau\int_{t_1}^{\tau} d\tau'\int_{t_1}^{\tau} d\tau_1 \frac{\delta^2 n_a(\tau)}{\delta u_a(\tau')\delta u_a(\tau_1)}
_{|_{u_a=0}}\nonumber\\
&+\dots,
\label{eq:model_cumulant_expansion}
\end{align}
{where $n_{a}^0$ is the occupation of the state $a$ in the ground-state, and the first term, linear in $U$ replaces the bare Coulomb interaction seen in Eq.\ (\ref{eq:final}) for the model Hamiltonian, and causes an overall quasiparticle shift. Note that there are also  further quasiparticle shifts that come from the higher order terms in Eq.\ (\ref{eq:model_cumulant_expansion}).}
]

Here, the density is given in terms of the time-dependent wavefunction, $n_a(t) = |\langle a|\psi(t)\rangle|^2$, where $|\psi(t)\rangle$ is the state of the system at time $t$ after the appearance of the core-hole, and is initially equal to the ground state valence wavefunction, $|\psi(0)\rangle=|\psi^{i}_0\rangle$.
This is the only component that contributes to the non-linear cumulant solution due to the fact that the core-level couples only to the atomic level $a$. 
As pointed out above, the density variations are linear and non-linear response functions, evaluated in the initial ground state, i.e. without a core hole. 
Finally we apply Eq.\ (\ref{eq:final}) for the cumulant, which yields
\begin{align}
&C(t_1t_2)= 
{iUn_a^0(t_2-t_1) + }iU\int_{t_1}^{t_2} d\tau \Delta \tilde n_a(\tau;[v_a]),\label{eq:model_cumulant_tdd}
\end{align}
where the time-dependent potential due to the core hole is $v_a(t)=0$ for $t<t_1$ and
$v_a(t)=U$ for $t>t_1$. We can now compare the results obtained from the two lowest orders of Eq.\ (\ref{eq:model_cumulant_expansion}) and the full  Eq.\ (\ref{eq:model_cumulant_tdd}). 


\begin{figure}
\centering
 \includegraphics[height=\columnwidth, angle=-90]{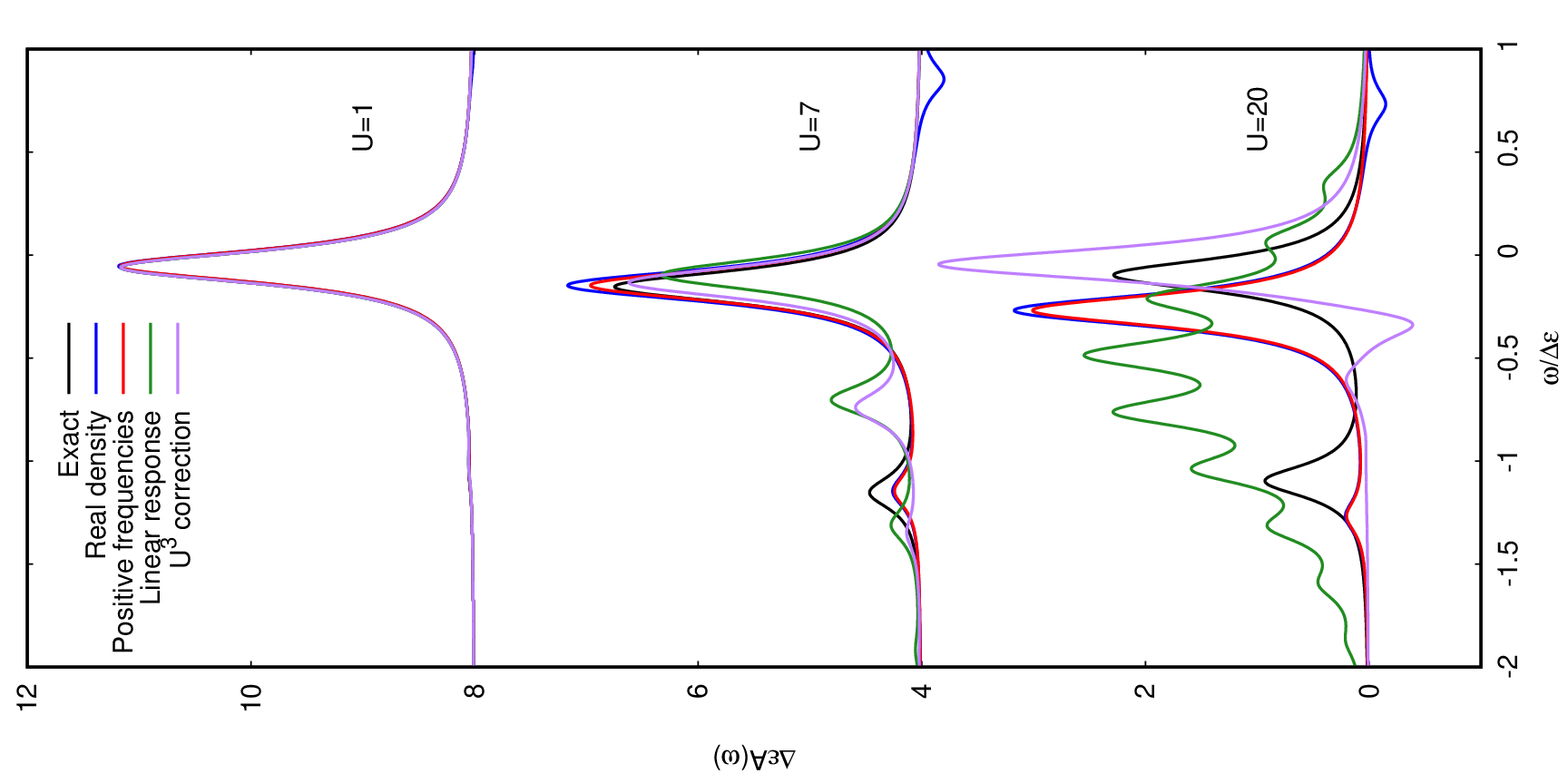}
 \caption{Normalized spectral function $A(\omega) = (-1/\pi) {\rm Im}\, G_c(\omega)$ vs frequency $\omega$  for model parameters $\epsilon$ = 1eV and $t$  = 3eV . The coupling $U = 1$ simulates the weak coupling to the core-hole (top), while $U = 7$ simulates intermediate coupling regime (middle), and $U=20$ strong coupling. Results are shown for the exact spectral function (black),  the real time-dependent density following Eq.\ (\ref{eq:final}) (blue), the same but with only positive frequencies included in the density $\Delta\tilde n$ (red), the linear response approximation (green), and from including the first non-linear ($U^3$) correction term (purple).
 Note that the results from the time-dependent density are in good agreement for the positions of the quasi-particle and satellite peaks for $U=1$ and $U=7$. Even at $U=20$ the excitation energy (difference between quasiparticle and satellite) matches that of the exact result, although the result from the real density produces small negative satellites on the high energy side of the quasiparticle. In contrast, the linear response and $U^3$ corrected approximations vastly under-estimate the satellite energy and give unphysical results, such as large boson-like satellite progression and negative spectral weight.}
 \label{fig:model}
\end{figure}

Results for the core-level spectral function $A_c(\omega)={-\frac{1}{\pi}\rm Im}\, G(\omega)$ vs. $\omega/\Delta\epsilon$, where $\Delta\epsilon 
{= \epsilon_{1}^f-\epsilon_{2}^f}$ is the valence excitation energy in the presence of the core-hole,  
are shown in Fig.\ (\ref{fig:model}). The position of the peaks reflects the energy of the transition of the valence electron from the initial (with the core electron) to the final (with the core-hole) state, while the height of the peaks corresponds to the probability amplitude of the transition. Only a non-negligible value of the parameter $t$ allows for the core-potential to affect the transition energies between the initial and the final state, since otherwise no screening can happen.
From top to bottom, the curves show results for core-hole strength $U=1$ (top), $U=7$ (middle), and $U=20$ (bottom), corresponding to weak, intermediate, and strong coupling, all with the parameters $t=3$, and $\epsilon = 1$. The top set of curves represent the weak coupling limit, which can be seen by the lack of any visible satellite, and by the agreement of all curves, which should be nearly identical in the linear response regime. At intermediate coupling ($U=7$, middle), the various approximations now give appreciably different results. The linear response approximation (green) underestimates the splitting between the satellite and quasiparticle positions by nearly a factor of two, underestimates the size of the quasiparticle peak, and produces a second satellite indicating a boson-like progression, as expected. The lowest order ($U^3$) corrected result (purple) gives a small correction to the quasiparticle weight and position, but does little to correct the splitting between the quasiparticle and satellite. The issues with the linear response and $U^3$ corrected approximations become even more apparent in the strong coupling regime ($U=20$), where the linear response vastly underestimates quasiparticle weight and quasiparticle-satellite splitting, and produces a long progression of satellites, initially with increasing weight
and overestimates the quasiparticle peak by a large amount. This is unphysical, since a single valence electron cannot produce multiple excitations. Also the inclusion of the $U^3$ correction produces 
{spurious} negative spectral weight. In contrast, the results of Eq.\ (\ref{eq:final}) with the real-time density used for $\Delta \tilde n(t)$ produces good agreement for the quasiparticle and satellite peak positions for $U=7$, and even reproduces the satellite splitting at $U=20$, although the satellite weight is underestimated in both cases, and the approximation produces negative spectral weight above the quasiparticle peak. To correct this unphysical behavior, we have used a form for $\Delta \tilde n(t)$ with the negative frequencies filtered out. This produces a Landau type form for the cumulant, similar to that of the linear response. Results for this approximation (red) show little difference in comparison to those obtained with the real density, apart from the lack of any negative spectral weight, showing that this approximation is a reasonable method for obtaining a physical result.

\section{Conclusions}
\label{sec:conclusions}

We have demonstrated that the Kadanoff-Baym functional differential equation is a convenient starting point to derive the form of the cumulant Green's function beyond the linear response approximation. For a single level that can be considered as decoupled from the rest of the system, such as a localized deep core level, the result can be formulated in a compact way, which highlights the essential physics: i.e., the sudden switching-on of a core hole perturbs the valence density, and the subsequent time-integral of the change in density leads to a quasi-particle correction and to satellites in the spectral function. The approach is tested on a simple 3-level model system similar to that of Lee, Gunnarsson and Hedin.  The numerical results suggest that for molecular systems with strong core-hole effects non-linear effects can  significantly improve the calculated photoemission spectra and generally need to be taken into account. We suggest that coupling this non-linear cumulant approach with real-time TDDFT can be a promising way to include  the non-linear effects in \textit{ab initio} calculations of the photoemission spectra for such systems.

 

\begin{acknowledgments} This work was carried out with support from the U.S. Department of Energy, Office of Science, BES, Chemical Sciences, Geosciences and Biosciences Division in the Center for Scalable and Predictive methods for Excitations and Correlated phenomena (SPEC) at PNNL, and with computer support from DOE-NERSC and from GENCI (544).
{LR,} MT and JR also acknowledge support at the beginning of this project from 
  the European  Research Council (ERC Grant Agreement No.~320971).
\end{acknowledgments}

\bibliographystyle{apsrev4-1} 
\bibliography{nonlin-bib}
\end{document}